\documentclass[aps,prb,showpacs,twocolumn,superscriptaddress]{revtex4-1}
\usepackage{amssymb}
\usepackage{graphicx}
\usepackage{appendix}
\usepackage{dcolumn}
\usepackage{bm}
\usepackage{amsmath}
\usepackage{epstopdf}
\usepackage{xcolor}
\usepackage{float}
\usepackage{braket}
\usepackage[colorlinks,urlcolor=blue,anchorcolor=blue,linkcolor=blue,citecolor=blue,breaklinks=true]{hyperref}
\usepackage[T1]{fontenc}
\usepackage{array}
\usepackage{booktabs}

\begin{document}
\title{Topological flat bands in hyperbolic lattices}
\author{Dong-Hao Guan}
\affiliation{National Laboratory of Solid State Microstructures and Department of Physics, Nanjing University, Nanjing 210093, China}
\affiliation{College of Physics Science and Technology, Yangzhou University, Yangzhou 225002, China}
\author{Lu Qi}
\affiliation{College of Physics Science and Technology, Yangzhou University, Yangzhou 225002, China}
\author{Yuan Zhou}
\email{zhouyuan@nju.edu.cn}
\affiliation{National Laboratory of Solid State Microstructures and Department of Physics, Nanjing University, Nanjing 210093, China}
\author{Ai-Lei He}
\email{heailei@yzu.edu.cn}
\affiliation{College of Physics Science and Technology, Yangzhou University, Yangzhou 225002, China}
\author{Yi-Fei Wang}
\affiliation{Zhejiang Institute of Photoelectronics $\&$ Zhejiang Institute for Advanced Light Source, Zhejiang Normal University, Jinhua 321004, China}
\affiliation{Center for Statistical and Theoretical Condensed Matter Physics, and Department of Physics, Zhejiang Normal University, Jinhua 321004, China}

\date{\today}

\begin{abstract}
Topological flat bands (TFBs) provide a promising platform to investigate intriguing fractionalization phenomena, such as the fractional Chern insulators (FCIs). Most of TFB models are established in two-dimensional Euclidean lattices with zero curvature. In this work, we systematically explore TFBs in a class of two-dimensional non-Euclidean lattices with constant negative curvature, {\emph i.e.,} the hyperbolic analogs of the kagome lattice. Based on the Abelian hyperbolic band theory, TFBs have been respectively found in the heptagon-kagome, the octagon-kagome, the nonagon-kagome and the decagon-kagome lattices by introducing staggered magnetic fluxes and the next nearest-neighbor hoppings. The flatness ratios of all hyperbolic TFB models are more than 15, which suggests that the hyperbolic FCIs can be realized in these TFB models. We further demonstrate the existence of a $\nu=1/2$ FCI state with open boundary conditions when hard-core bosons fill into these hyperbolic TFB models.
\end{abstract}
\maketitle

{\noindent {\it Introduction.---}}Chern insulators (CIs) or quantum anomalous Hall states, the realization of quantum Hall states in the absence of external magnetic field, have stimulated wide interests theoretically and experimentally. For decades several CI models have been sequentially proposed in periodic
lattices~\cite{CB0,KG0,LDM,Star0,KG1,Star1,SQOC0,Lieb0,KG2,CB1,Neupert,Lieb1,Ruby,Star2,TR,dice2,Lieb2,KG3, SQOC1,2DSSH_flux,SQOC3} with the aid of staggered magnetic fluxes. The study of CIs has been also extended to two-dimensional (2D) singular lattices, quasicrystals and amorphous systems~\cite{QCCI,Amorphous0,Amorphous1}. These CI models host non-trivial energy bands with non-zero Chern numbers and satisfy the bulk-edge correspondence. Notably, quantum anomalous Hall states have been observed in the magnetic topological insulator~\cite{RealizeCI01,RealizeCI02} and the twisted bilayer systems~\cite{RealizeCI03,RealizeCI04} in recent experiments. Ultracold fermionic systems~\cite{RealizeCI1}, acoustic systems~\cite{RealizeCI2}, photonic crystals~\cite{RealizeCI3} and superconducting quantum processors~\cite{RealizeCI4} also offer the possibility of realizing CIs. Most recently, the concept of CIs has been extended into hyperbolic lattices~\cite{CIhyper0,CIhyper1,CIhyper2,CIhyper3} which is the uniform discretizations of 2D hyperbolic geometry with constant negative curvature.

Bands of CIs can be flattened by tuning the hopping parameters theoretically and a series of topological flat band (TFB) models based upon 2D Euclidean lattices have been proposed~\cite{CB1,KG2,Neupert,KG3,TR,Ruby,Star2}. Since these TFBs can quench the kinetic energy and enhance the particle-particle interaction effectively, they play a crucial role in the realization of fractional Chern insulators (FCIs)~\cite{CB1,KG2,Neupert}. Several analytical and numerical works have systematically explored the interaction effects within the TFB models, and convincing evidences have revealed the existence of FCIs~\cite{Sheng1,Regnault1,WYFTFB,YFWang2,Qi0,GPPc,YLWu,YLWu0,Ronny0,Ronny1,Ronny3,LiuTH,LiuZ0,LiuZ2,LiuZ3,YLWu3,WWLuo,HeAL21,FCI_reviews,FCI_reviews1,FCI_reviews2}. Most significantly, FCIs have been theoretically explored in Moir{\'e} superlattice systems~\cite{MFCI01,MFCI02,MFCI03,MFCI04,MFCI1,MFCI2,MFCI3,MFCI4} and related experimental signatures have been recently reported~\cite{Cai2023,Zeng2023,Park2023,XuFan2023,lu2023fractional}. Additionally, FCIs have been studied in singular TFB models established in singular geometries with non-zero Gauss curvature around the singularity~\cite{HeAL3,HeAL5}. Geometric factor of these singular lattices, as an implicit degree of freedom, plays a crucial role in these FCIs~\cite{HeAL3,HeAL5}. And multiple branches of edge excitations (EEs) emerge in this FCI state because of the defect-core states~\cite{HeAL5}.

\begin{figure}
\includegraphics[scale=0.4]{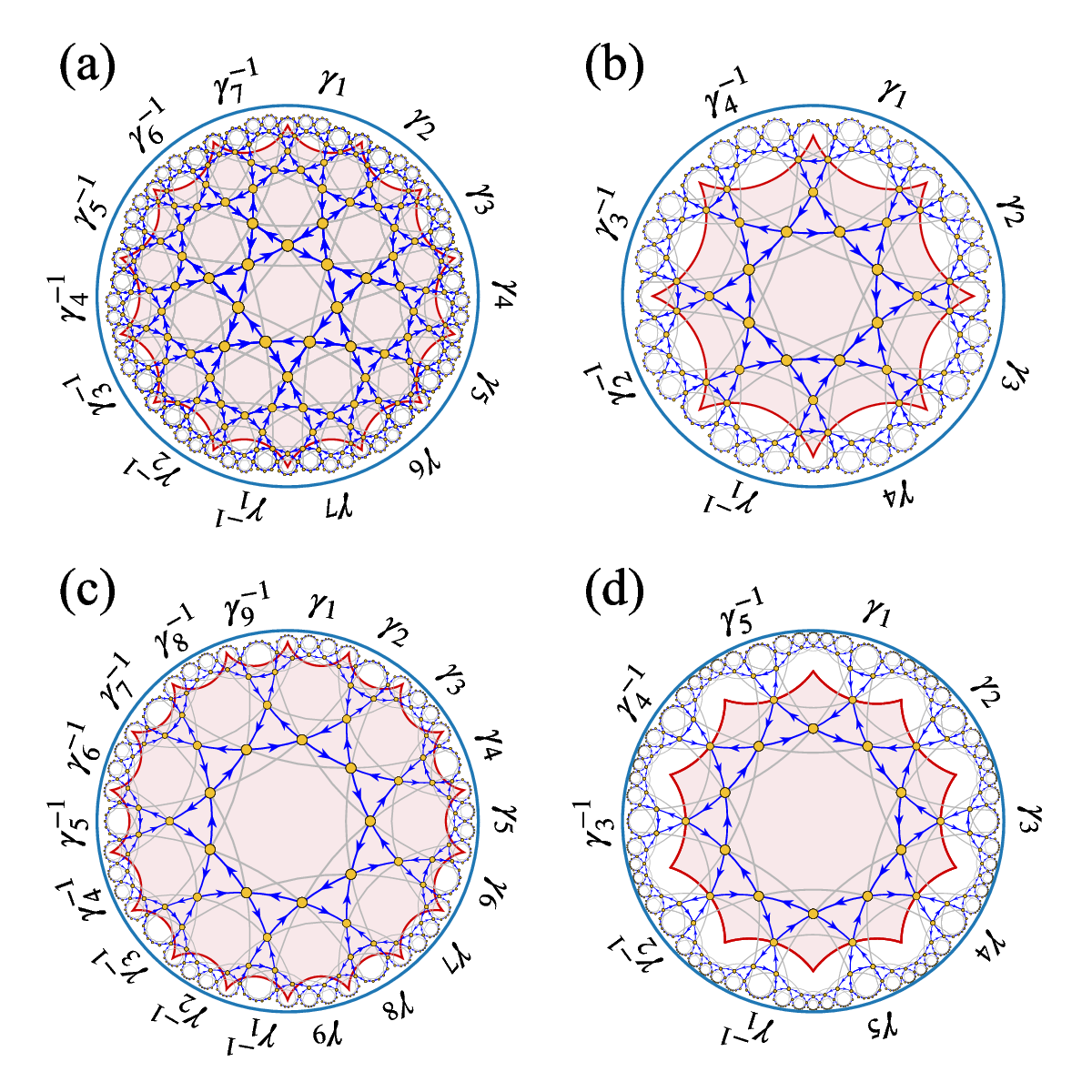}
\caption{(Color online) Schematic structure of hyperbolic analogs of the kagome-lattice model with NN and NNN hoppings. Depending on the geometrical conditions, the unit cells (within red shades) respectively contain 84, 24, 54, and 15 atoms for the (a) HKG, (b) OKG, (c) NKG and (d) DKG models. Blue and gray bonds denote the NN and NNN hoppings with the corresponding hopping potentials $t_1$ and $t_2$. The staggered magnetic fluxes induced additional phase factors $\phi$ on the NN bonds are denoted by the arrows. $\gamma _i$ and $\gamma _{i}^{-1}$ are the generators of translation in hyperbolic space where the number of generators is equal to twice the number of genus as shown in TABLE~\ref{tab1}}
\label{Model}
\end{figure}

Unlike singular lattices, 2D hyperbolic lattices are with constant negative curvature, but without any singularities. In stark contrast to 2D Euclidean lattice models, hyperbolic lattice models feature hyperbolic band theory and crystallography~\cite{Joseph,Joseph1,Boettcher,ChengN,Kazuki,Elliot,Patrick}, periodic boundary conditions and  thermodynamic limit~\cite{Lux,Mosseri,NobleG}, simulation of holographic duality~\cite{Basteiro} and conformal field theories~\cite{CFTs}, unique Anderson localization transitions~\cite{HAnder1,HAnder2,HAnder3}, etc. Additionally, topological states~\cite{CIhyper0,CIhyper1,CIhyper2,Kazuki2,YuS,Stegmaier,Zhang2022,Zhang2023,LiuZR1,PeiQS,TaoYL,TarunT}, quantum spin liquids~\cite{SLHyper,CSLHyper} and non-Hermitian physics~\cite{SunJ} have been investigated in hyperbolic systems. More interestingly, hyperbolic lattices can host unusual flat bands~\cite{FBH1,FBH2} and TFBs~\cite{TFBZhang,HFCIs}. Besides the theoretical studies, there are also several experimental schemes proposed to realize the hyperbolic lattice structures based on circuit quantum electrodynamics~\cite{Kollar2019}, classical electric-circuit networks~\cite{Lenggenhager2022,Chen2023,ReahyCI1,ReahyCI2,CIhyper3}, microwave platforms~\cite{ReahyCI3} and photonic systems~\cite{ReahyCI4}. Most recently, hyperbolic TFBs and FCIs~\cite{HFCIs} have been explored in several hyperbolic analogs of the kagome-lattice model with open boundary conditions. However, the flatness ratios of these hyperbolic TFBs are not directly measured and the best hyperbolic TFBs are barely found in the previous work~\cite{HFCIs}.

In this work, we systematically investigate the TFBs in hyperbolic analogs of the kagome lattice with periodic boundary condition based on the Abelian hyperbolic band theory. We specifically focus on four hyperbolic analogs of the kagome lattice: the heptagon-kagome (HKG), the octagon-kagome (OKG), the nonagon-kagome (NKG) and the decagon-kagome (DKG) lattices. By introducing the next nearest-neighbor (NNN) hoppings and staggered magnetic fluxes along the nearest-neighbor (NN) hoppings, TFBs have been respectively found in these hyperbolic lattices. Each model hosts TFBs with flatness ratio more than 15 which can be candidates to investigate FCIs. When hard-core bosons fill into these hyperbolic TFB models, $\nu=1/2$ FCI state emerges with open boundary condition which is characterized by the angular momentum of ground state and the quasi-degenerate sequence of their EEs. Intriguingly, there exists multiple branches of EEs because of the central localized state which has been explained in the recent work~\cite{HFCIs}.

{\noindent {\it Models.---}}The hyperbolic kagome lattices originate from the $\left\{ p,q \right\}$ hyperbolic lattices, where $q$ represents the number of regular $p\mbox{-}gons$ which a vertex is shared by. If triangles are substituted for the vertices in a $\left\{ p,q \right\}$ hyperbolic lattice, the hyperbolic analogs of the kagome lattice can be obtained. For example, the HKG lattice can be achieved by replacing the vertices in a $\{7,3\}$ lattice with triangles. Other hyperbolic analogs of the kagome lattice (such as the OKG, NKG and DKG) can be obtained and they can be represented by using the Poincar{\'e}-disk model (details shown in Fig.~\ref{Model}).

Here, we specifically explore TFBs in the HKG, OKG, NKG and DKG models with periodic boundary conditions. Based on the crystallography of hyperbolic lattices~\cite{Boettcher}, the associated unit cells of these hyperbolic kagome lattices can be obtained (see Fig.~\ref{Model}). These 2D hyperbolic lattices exist the translational symmetry and the translational directions are labeled by $\gamma_1, \gamma_2, \gamma_3,...$ and $\gamma^{-1}_1,\gamma^{-1}_2,\gamma^{-1}_3,...$ (details in Fig.~\ref{Model}). Accordingly, the corresponding Brillouin zones (BZs) have more than two dimensions. For instance, there is a 14-sided unit cell with $V_0=56$ vertices in the $\{7,3\}$ hyperbolic lattice~\cite{Boettcher} and the corresponding HKG lattice contains $V=84$ vertices in the unit cell. The translational directions are $\gamma_1, \gamma_2, \gamma_3,...,\gamma_7$, however, there are only six independent components of hyperbolic crystal momentum {\emph i.e.}, ${\bf k}=\{k_1,k_2,k_3,k_4,k_5,k_6\}$ because of the constraint on the momentum $k_1+k_2+k_3+k_4+k_5+k_6+k_7=0$~\cite{Boettcher}. Notably, the HKG model hosts 6D BZ which corresponds to a closed manifold of genus $g=3$, in stark contrast to the 2D BZ which corresponds to a torus with genus $g=1$. Similarly, the number of vertices in a unit cell and the BZ manifolds of other hyperbolic kagome-like lattices can be obtained with the aid of the crystallography of hyperbolic lattices~\cite{Boettcher} (details in Table~\ref{Model}). Additionally, differing from the 2D Euclidean lattice models whose translation groups are Abelian, the translation groups of hyperbolic lattices are the Fuchsian groups, which belong to non-Abelian group. In this work, we only consider the Abelian hyperbolic band theory and the corresponding Abelian Bloch state $\psi_{\bf k}(z)$ satisfies $\psi_{\bf k}( \gamma_j (z))=e^{ik_j}\psi_{\bf k}(z)$ where $\gamma_j$ is a generator of the non-Abelian translation group.

\begin{table}[ht]
\center
\caption{List of four hyperbolic kagome lattices with regular Bravais lattices of genus $g$. The number of atoms in a unit cell is denoted by $V$. $V_0$ marks the number of unit cell sites of regular $\{p,q\}$ hyperbolic lattices. The dimension of the higher-dimensional Brillouin zone (BZ) for each model is present.}
\begin{tabular}{m{40pt}<{\raggedright}m{40pt}<{\raggedright}m{40pt}<{\raggedright}m{40pt}<{\raggedright}m{40pt}<{\raggedright}}
\hline\\[-2.9mm]\hline
  Model &  $V_0$ &  $V$ &$g$ &  ${\rm BZ}$  \\
\hline
  HKG   &   56    &   84     &3    & 6D       \\
  OKG   &   16    &   24     &2    & 4D      \\
  NKG   &   36    &   54     &4    & 8D       \\
  DKG   &   10    &   15     &2    & 4D       \\
\hline\\[-2.9mm]\hline
\end{tabular}\label{tab1}
\end{table}

Inspired by TFBs in the 2D Euclidean kagome lattice, we introduce the NN and NNN hoppings in these hyperbolic kagome models and the staggered magnetic fluxes are added along the NN hoppings. The corresponding tight-binding Hamiltonian can be written as
\begin{eqnarray}\label{HAMR}
 \begin{aligned}
H & = t_1\sum_{\left. \left< \boldsymbol{r}^{\prime}\boldsymbol{r} \right> \right.}{\left( a_{\boldsymbol{r}^{\prime}}^{\dagger}a_{\boldsymbol{r}}e^{i\phi _{\boldsymbol{r}^{\prime}\boldsymbol{r}}}+\mathrm{H}.\mathrm{c}. \right)}\\
&+ t_2\sum_{\left. \left< \left< \boldsymbol{r}^{\prime}\boldsymbol{r} \right> \right> \right.}{\left( a_{\boldsymbol{r}^{\prime}}^{\dagger}a_{\boldsymbol{r}}+\mathrm{H}.\mathrm{c}. \right)}
\\
\end{aligned}
\end{eqnarray}
where $a_{\mathbf{r}}^{\dagger}(a_{\mathbf{r}})$ creates (annihilates) a spinless particle at $\mathbf{r}$. $\left \langle ...\right \rangle$ and $\left \langle \left \langle ... \right \rangle \right \rangle $ denote the NN and NNN pairs of sites and the hopping integrals are $t_{1}$, $t_{2}$, respectively. $\phi _{\boldsymbol{r}^{\prime}\boldsymbol{r}}=\pm \phi$ is the phase factor along with the NN hoppings, and the sign $\left( \pm \right)$ is denoted by the direction of arrows in Fig.~\ref{Model}. One can easily find that the total fluxes of these hyperbolic lattice models are vanished. For simplicity, we set the NN hopping integral as $t=-1.0$.

\begin{figure}[htp]
\includegraphics[scale=0.4]{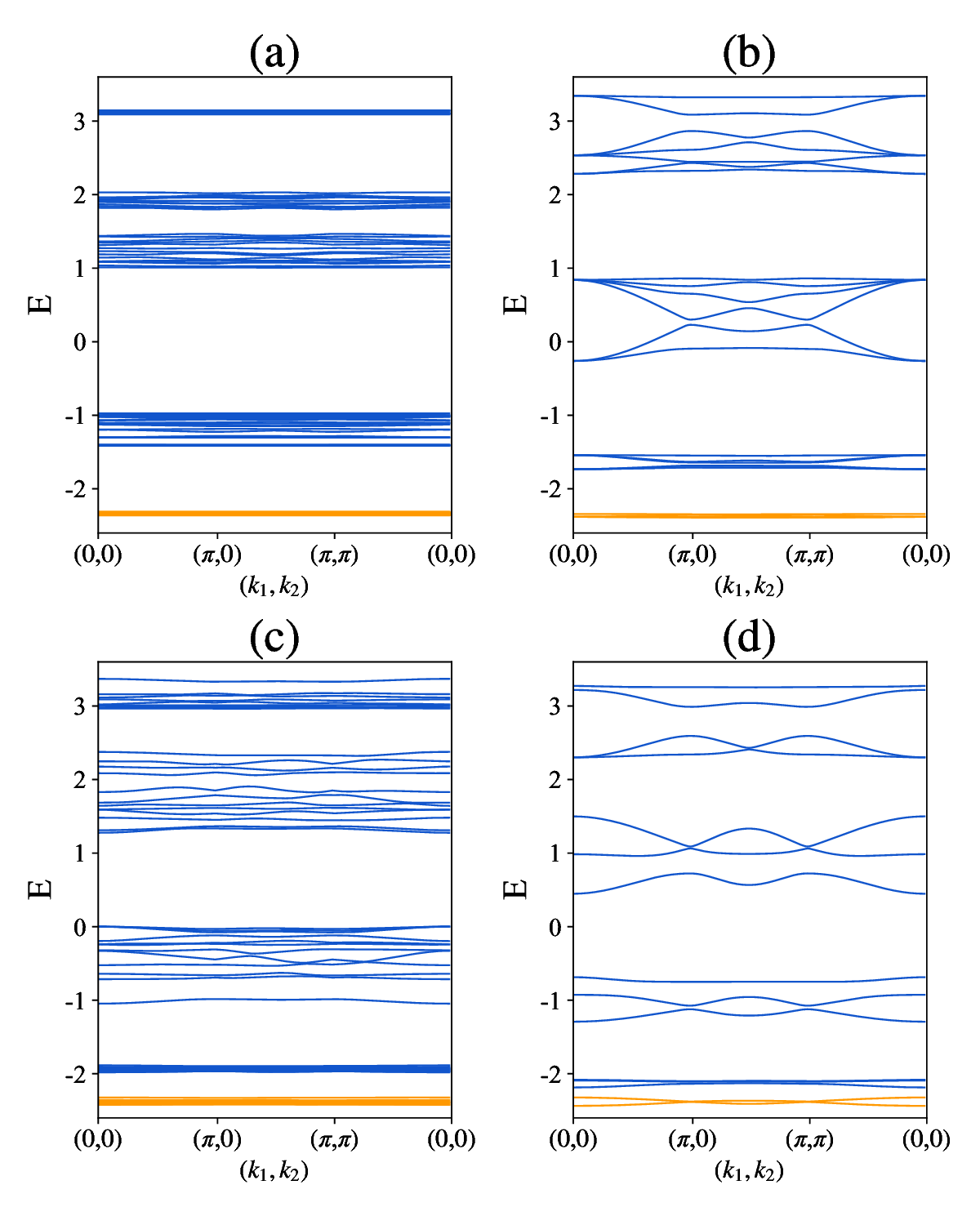}
\caption{(Color online) Bulk energy bands for (a) HKG, (b) OKG, (c) NKH and (d) DKG models with $t_1=-1$, $t_2=0.19$ and $\phi=0.22\pi$. Here, we consider a 2D sub-torus spanned by $(k_1,k_2)$ and other momenta are zero. The lowest flat subbands are colored with orange.}
\label{TFBs}
\end{figure}

\begin{figure}[htp]
\includegraphics[scale=0.42]{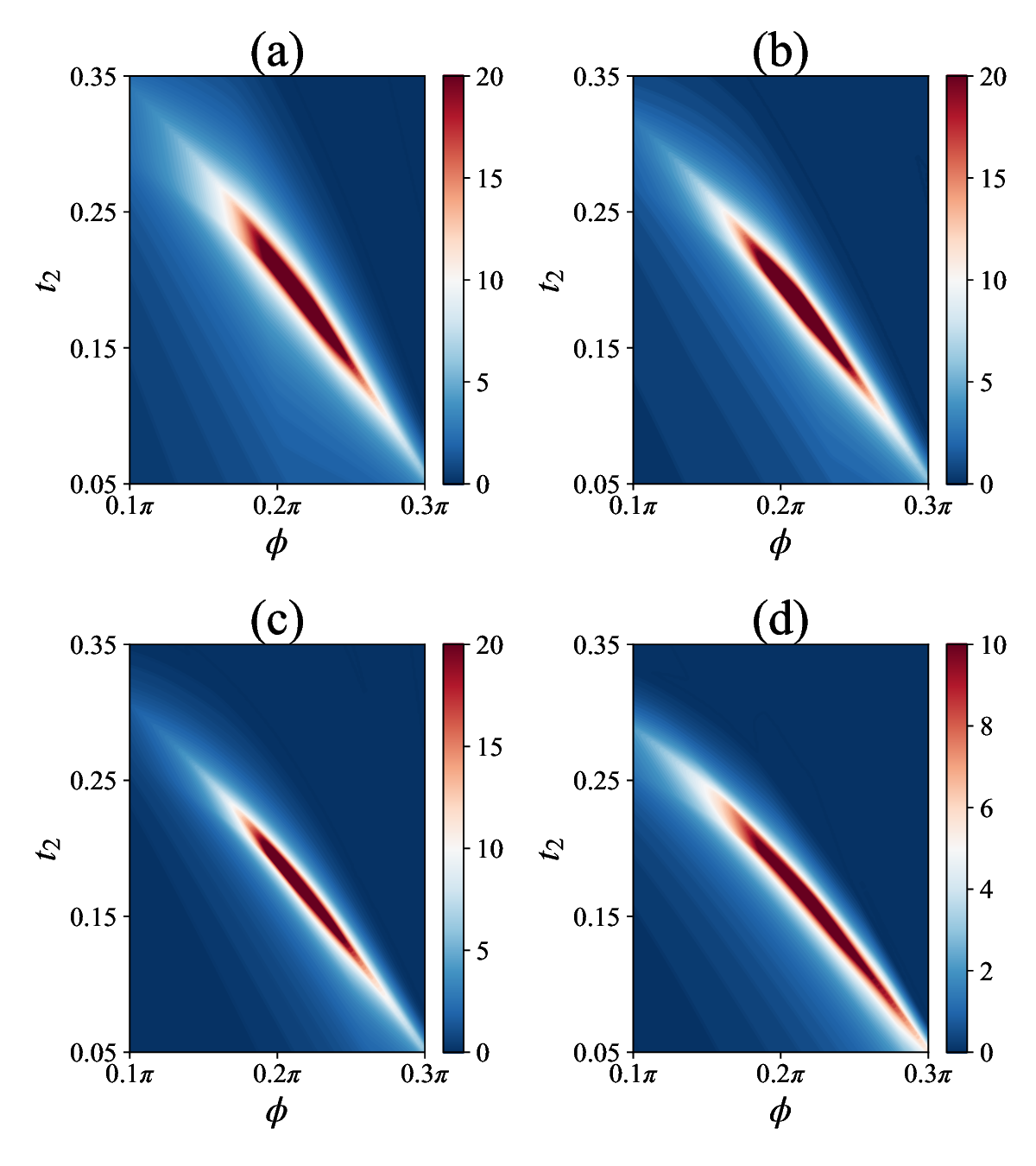}
\caption{(Color online) Color map of the flatness ratio $\epsilon$ as a function of the NNN hopping potential $t_2$ and phase factor $\phi$ for (a) HKG, (b) OKG, (c) NKG and (d) DKG models. Here, we set the range of NNN hopping integral $t_2 \in [0.05,0.35]$ and phase factor $\phi \in [0.1\pi,0.3\pi]$. Color bars denote the size of flatness ratio.}
\label{TFBs1}
\end{figure}

{\noindent {\it Topological flat bands.---}}TFBs in 2D Euclidean kagome~\cite{KG2,KG3} and OKG~\cite{TFBZhang} lattice models have been found with suitable hopping parameters (details in the Supplemental Material~\cite{SM}). Here, we systematically explore TFBs in several hyperbolic kagome models on the basis of 2D Euclidean kagome TFB with hopping parameter $t=-1.0$, $t_2=0.19$ and $\phi=0.22\pi$~\cite{KG3}. For a 2D Euclidean kagome CI, one can define the flatness ratio to measure the flatness of its lowest energy band, {\emph i.e.}, $\epsilon_1=\Delta_1/W_1$, where $\Delta_1$ is the band gap between the lowest band and the second lowest band, and $W_{1}$ is the band width of the lowest band. However, there are multiple low-energy subbands mixing together in the OKG model~\cite{TFBZhang} as well as in other kagome-like models (see Fig.~\ref{TFBs}). Accordingly, the flatness ratio is defined as $\epsilon = \Delta/W$ where $W$ is the bandwidth of the low-energy subbands and $\Delta$ is the size of gap between the mixing low-energy subbands and their adjacent high-energy band.

Based on the Abelian hyperbolic band theory~\cite{Joseph,Joseph1} and crystallography~\cite{Boettcher} of hyperbolic kagome lattices, we can obtain the energy bands by diagonalizing Hamiltonian in high dimensional BZs. We first consider the hyperbolic kagome models with the TFB hopping parameters of 2D Euclidean kagome~\cite{KG3}, and present the bulk bands on a 2D sub-torus spanned by two momenta $(k_1,k_2)$ with other momenta zero (details as displayed in Fig.~\ref{TFBs}). Notably, there are multiple low-energy subbands mixed together, for example, 22 low-energy subbands in the flat band for HKG model. We can calculate the flatness ratio of these multiple low-energy subbands in the whole BZs and based on the definition of flatness ratio $\epsilon$, we obtain $\epsilon =18, 10, 3$ and $1$ for the HKG, OKG, NKG and DKG models. For the HKG and OKG models, the flatness ratios are more than 10 which suggests the existence of TFBs in these two models. However, the flatness ratios are not very high for the remaining two models. We therefore search TFBs with higher flatness ratio for these hyperbolic kagome models in the $\{t_2,\phi\}$ parameter space (the corresponding results shown in Fig.~\ref{TFBs1}).

\begin{figure}[htp]
\includegraphics[scale=0.42]{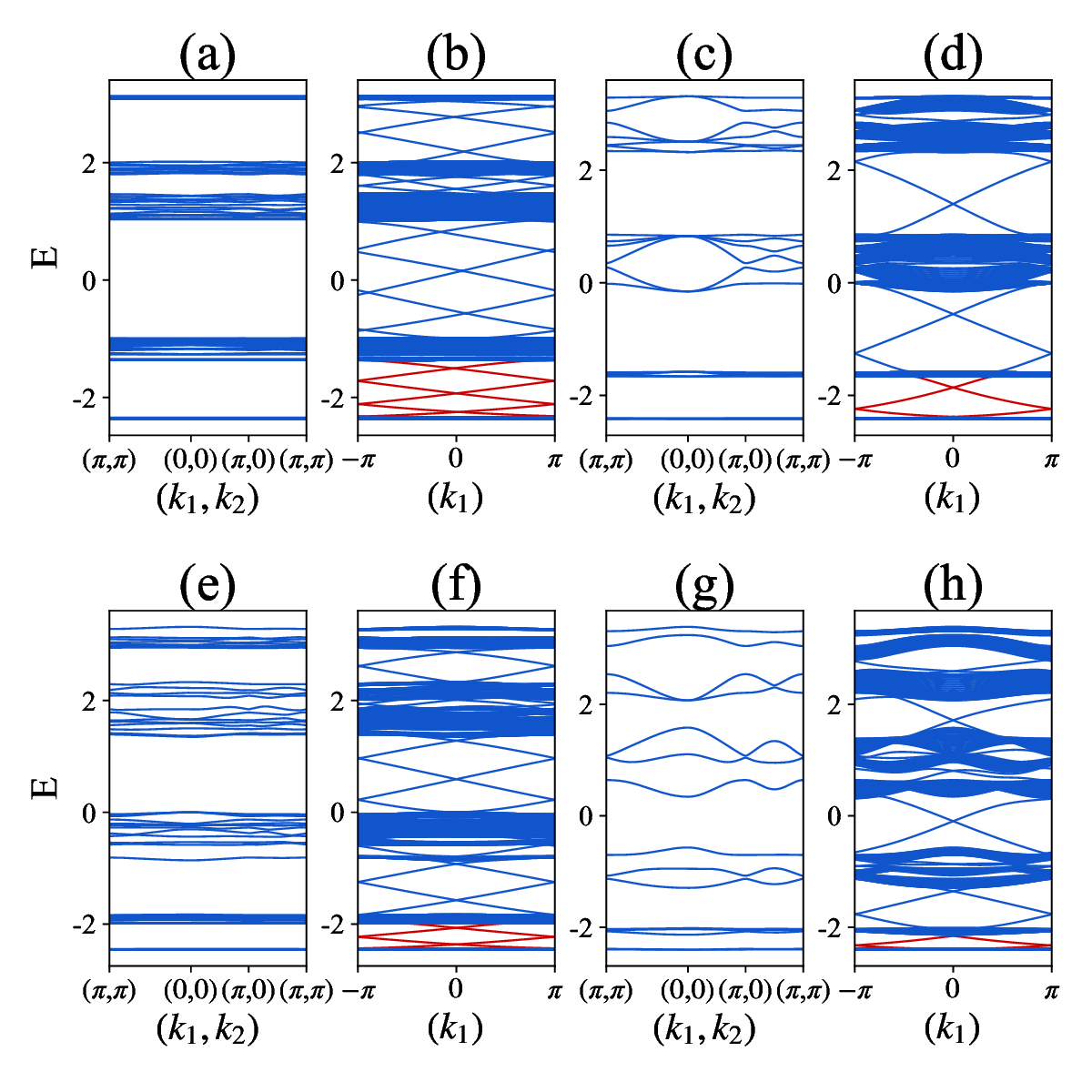}
\caption{(Color online) Energy bands of hyperbolic TFBs. (a), (c), (e) and (g) are the bulk energy bands in the ($k_1$,$k_2$) subspaces of the HKG, OKG, NKH and DKG models. (b), (d), (f) and (h) are the energy bands with periodic boundary condition along $\gamma_1$ and open boundary along $\gamma_2$. The lowest-energy boundary states are colored with red. Other momenta are zero.}
\label{Latts0}
\end{figure}

The best flat bands with flatness ratios 35, 45, 30, and 15 are found for these hyperbolic kagome models (see Fig.~\ref{TFBs1} and the corresponding hopping parameters in the Supplemental Material~\cite{SM}) and their bulk bands are shown in Fig.~\ref{Latts0}(a), \ref{Latts0}(c), \ref{Latts0}(e) and \ref{Latts0}(g) on a 2D sub-torus spanned by two momenta $(k_1,k_2)$. To identify the topological properties of these flat bands, we first calculate the Chern number matrix based on the definition of Chern number in 2D reciprocal spaces ($\emph i.e.$, on a 2D sub-torus spanned by two momenta). We find the off-diagonal elements of Chern number matrix are nonzero and equal to one or minus one (details shown in the Supplemental
Material~\cite{SM}). Additionally, edge states can be used to characterize the topological properties. Here, we consider theses hyperbolic models on a 1D sub-cylinder with periodic boundary condition along $\gamma_1$ (corresponding to $k_1$), open boundary condition along $\gamma_2$ and the remaining momenta are zero. With this special periodic boundary condition, we can obtain chiral edge states connecting the valence and conduction bands [Fig.~\ref{Latts0}(b), \ref{Latts0}(d), \ref{Latts0}(f) and \ref{Latts0}(h)], which reveals the bulk-edge correspondence in hyperbolic kagome models. Accordingly, based on the chiral edge states and Chern number matrix, we demonstrate the existence of hyperbolic TFBs for these four hyperbolic kagome models.

The TFB has been reported in the OKG model~\cite{TFBZhang}, however, numbers of mixing bulk subbands in these two TFBs are significantly different.  There are three bulk subbands mixing in the TFB~\cite{TFBZhang} instead of five subbands in the present OKG model, and the flatness ratios of these two TFBs exist differences (details in the Supplemental Material~\cite{SM}). Especially, except for the OKG model, we find TFBs can be achieved in other hyperbolic kagome models. In addition, TFBs of the hyperbolic kagome models with open boundary conditions are also obtained when consider the similar TFB parameters with periodic boundary conditions (detains in the Supplemental
Material~\cite{SM}). These hyperbolic TFBs are very robust against the size effect. And the non-Abelian translation of hyperbolic kagome models seems to have a negligible impact on the hyperbolic TFBs because these TFBs with periodic boundary conditions are almost consistent with the ones approximating the thermodynamic limit (see the Supplemental
Material~\cite{SM}).

\begin{figure}[htp]
\includegraphics[scale=0.83]{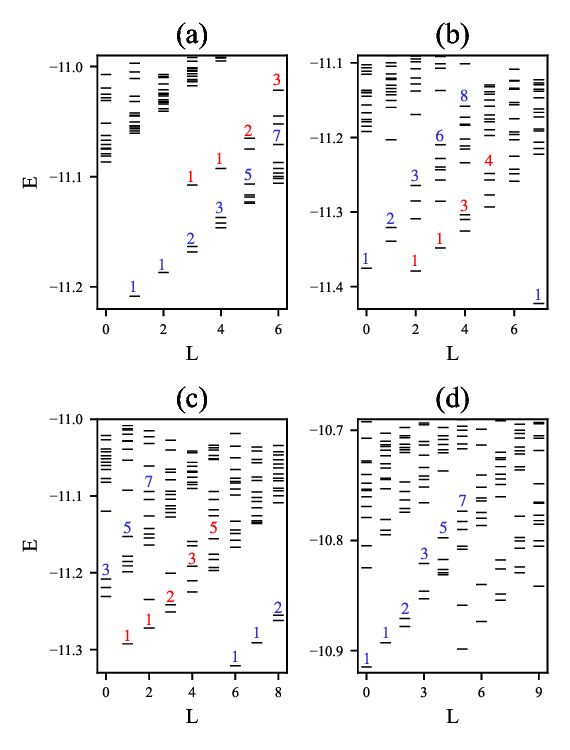}
\caption{(Color online) Edge excitations for $\nu=1/2$ FCI in (a) 98-site HKG, (b) 104-site OKG, (c) 108-site NKG and (d) 120-site DKG TFB models with trap potential $V_{\rm trap}=0.02$. Here, we consider five hard-core bosons loaded into these TFB models. No interaction is added for the HKG model, {\emph i.e.}, $V_{\rm NN}=0$. Interactions along NN bonds are added and $V_{\rm NN}=2$ for the OKG model. Interactions along NN bonds are considered ($V_{\rm NN}=V_{\rm NNN}=2$) for the NKG model, and $V_{\rm NN}=V_{\rm NNN}=20$ for the DKG model. Here, $V_{\rm NN}$ and $V_{\rm NNN}$ denote the NN and NNN interaction strengths.}
\label{EEs}
\end{figure}

{\noindent {\it Fractional Chern insulators.---}} FCIs have been explored in 2D Euclidean TFB models with various boundary conditions and the study of FCIs have been extended to singular TFB models which hosts non-zero Gauss curvature. Here, we use  the exact diagonalization (ED) method to explore hyperbolic FCIs in disk geometry when hard-core bosons fill into these hyperbolic TFB models. To confine the FCI droplet and make the edge modes propagate around the boundary, we add an additional trap potential in the hyperbolic disks. In general, the harmonic trap potential with the form $V=V_{\rm trap} \sum_{\bf {r}} |{\bf r}|^2 n_{\bf r}$ is adopted~\cite{EEkjall,EEWWL}. Here $V_{\rm trap}$ is the strength of trap potential and $n_{\bf r}$ is the number of occupied particles. $|{\bf r}|$ is the radius from the hyperbolic disk center and $|{\bf r}|={arcosh} [ (1+|z|^2) / (1-|z|^2) ]$ where $z=x+iy$ is the complex coordinate in the Poincar{\'e} disk.

We consider hard-core bosons loaded into these hyperbolic TFB models. According to the ED results, we find clear EE spectra for these hyperbolic TFB models with multiple branches of EEs (Fig.~\ref{EEs}) because of the existence of the ``central-localized'' orbital which leads to multiple occupation configurations for interacting particles~\cite{HFCIs}. The quasi-degenerate sequence of each branch is ``1,1,2,3,5,...'' which is reminiscent of Laughlin-type FCIs. To demonstrate the existence of $\nu=1/2$ bosonic FCI, we check the total angular momentum of the ground state. We first take the many-body state in HKG model with five bosons as an example and the corresponding root configuration of $\nu=1/2$ FCI is ``$|1010101010\rangle$''. We can predict the angular momentum based on the generalized Pauli principle~\cite{GPP1,GPP2,GPP3}. The corresponding angular momentum of ground state is $(6+1+3+5+0)\ {\bf mod} \ 7 = 1$ for a $\nu=1/2$ FCI in HKG model with five bosons, which is in accordance with the ED results [Fig.~\ref{EEs}(a)]. $\nu=1/2$ FCI can be obtained in the OKG, NKG and DKG TFB models when considering interacting hard-core bosons loaded into TFBs which can be substantially characterized by the EEs [Fig.~\ref{EEs}(b)-(c)]. These many-body states can be convincingly identified as a $\nu=1/2$ FCI state based on the trial wave functions~\cite{HFCIs}. Additionally, contrary to a $\nu=1/2$ FCI in HKG, OKG and NKG TFB models, the present $\nu=1/2$ FCI in DKG TFB model belongs to unconventional FCI which one particle occupies into the ``central-localized'' orbital~\cite{HFCIs} with angular momentum eight and the total angular momentum of the ground state is $(8+0+2+4+6)\ {\bf mod} \ 10 = 0$. Even though we add extremely strong repulsive interactions along the NN and NNN bonds ($V_{\rm NN}=V_{\rm NNN}=20$), the ground state still belongs to the unconventional FCI because the ``central-localized'' orbital closely mix into the TFB orbitals [Fig.~\ref{EEs}(d)].

{\noindent {\it Summary.---}} We explicitly demonstrate the existence of TFBs in hyperbolic analogs of the kagome lattice with staggered magnetic fluxes. Based on the Abelian hyperbolic band theory, the band structures can be obtained by diagonalizing Hamiltonian with periodic boundary conditions. By tuning hopping parameters, the flatness ratio of TFBs in these hyperbolic kagome models are more than 15 and the topological properties of these flat bands can be characterized by edge states and Chern number matrix. Subsequently, we explore FCI states with open boundary conditions by considering hard-core bosons filling into these TFB models and there are multiple branches of EEs based on the ED results. These FCI states are identified as a $\nu=1/2$ FCI state on account of the degeneracy sequence and the angular momentum of the ground state. Our study not only reveal the existence of hyperbolic FCIs, but also provide a systematic approach to find more hyperbolic TFBs.

{ {\it Acknowledgments}}---
This work was supported in part by the NSFC under Grants Nos. 12204404 (A.-L.H.),  12304557 (L.Q.) and 11874325 (Y.-F.W.) and the Natural Science Foundation of Jiangsu Province Grant No. BK20231397 (Y.Z.).

\bibliography{HyperTFBs}
\end{document}